\documentclass{jnmp}

\usepackage{amsmath}

\JNMPnumberwithin{equation}{section}

\newtheorem*{statement}{Statement}

\begin{document}

\renewcommand{\evenhead}{E~Poletaeva}
\renewcommand{\oddhead}{Superconformal Algebras
and Lie Superalgebras of the Hodge Theory}

\thispagestyle{empty}

\FirstPageHead{10}{2}{2003}{\pageref{poletaeva-firstpage}--\pageref{poletaeva-lastpage}}{Letter}

\copyrightnote{2003}{E~Poletaeva}

\Name{Superconformal Algebras\\
and Lie Superalgebras of the Hodge Theory}
\label{poletaeva-firstpage}

\Author{E~POLETAEVA}

\Address{Centre for Mathematical Sciences,
Lund University, Box 118, S-221 00 Lund, Sweden\\
E-mail: elena@maths.lth.se}

\Date{Received August 14, 2002; Revised September 19, 2002;
Accepted October 16, 2002}

\begin{abstract}
\noindent
We observe a correspondence between the zero modes of
superconformal
algebras $S'(2, 1)$ and $W(4)$ ([8]) and the Lie superalgebras
 formed by  classical operators appearing in the K\"ahler and
 hyper-K\"ahler geometry.
\end{abstract}

\section{Lie superalgebras of the Hodge theory}

\subsection{K\"ahler manifolds}
Let $M=(M^{2n}, g, I,\omega)$ be a compact
K\"ahler manifold of real dimension $2n$, where $g$ is
a~Riemannian (K\"ahler) metric,
$I$ is a complex structure on $M$, and
$\omega$ is the corresponding  closed 2-form defined by
$\omega (x, y) = g(x, I(y))$ for any vector fields $x$ and
$y$~[10].

A number of  classical operators on the Dolbeault algebra
$A^{*,*}(M)$
of complex diffe\-ren\-tial forms on $M$ is well-known~[5]:
the exterior differential $d$ and its holomorphic and
antiholomorphic parts,
 $\partial$ and $\bar{\partial}$, and $d_c= i(\partial -
\bar{\partial})$,
their dual operators, and the associated
Laplacians. Recall that
\begin{gather}
\partial: A^{p,q}(M) \rightarrow A^{p+1,q}(M),\qquad
\bar{\partial}: A^{p,q}(M) \rightarrow A^{p,q+1}(M),
\qquad d = \partial + \bar{\partial}.
\end{gather}
The Hodge operator $\star : A^{p,q}(M)\longrightarrow
A^{n-q,n-p}(M)$,
satisfies $\star^2 = (-1)^{p+q}$ on  $A^{p,q}(M)$.
Accordingly, the Hodge inner product is defined on  each of
$A^{p,q}(M)$:
$(\varphi, \psi) = \int_{M}\varphi\wedge \star\bar{\psi}$.
Recall that
$\triangle = dd^* + d^*d = 2\triangle_{\partial} =
2\triangle_{\bar{\partial}}.$
In addition,
$A^{*,*}(M)$ admits an $\mathfrak{sl}(2)$-module structure, where
$\mathfrak{sl}(2) = \langle E, H, F \rangle$ and the generators
satisfy
\begin{equation}
[E, F] = H, \qquad [H, E] = 2E, \qquad [H, F] = -2F.
\end{equation}
The operator
$E : A^{p,q}(M)\rightarrow A^{p+1,q+1}(M)$
is defined by
$E (\varphi) = \varphi \wedge \omega$.
(Clearly, $\omega$~is a~$(1, 1)$-form).
Let $F = E^*: A^{p,q}(M)\rightarrow A^{p-1,q-1}(M)$ be its dual
operator, and
$H\vert_{A^{p,q}(M)} = p + q - n.$
According to the Lefschetz theorem, there exists the
correspon\-ding
action of $\mathfrak{sl}(2)$ on $H^*(M)$~[5].
These operators satisfy a series of identities,
known as the {\it Hodge identities}~[5].
Let $\mathcal K$ be the Lie superalgebra, whose even part is
spanned by
$\mathfrak{sl}(2) = \langle E, H, F \rangle$
and the Laplace operator $\triangle$, and the odd part
is spanned by the differentials $d$, $d^*$, $d_c$, $d_c^*$.
The non-vanishing commutation relations in $\mathcal K$ are (1.2)
and the following relations (see~[5]):
\begin{gather}
[d, d^*] = [d_c, d_c^*] = \triangle,\nonumber\\
[H, d] = d, \qquad [H, d^*] = -d^*, \qquad [H, d_c] = d_c, \qquad
[H, d_c^*] = - d_c^*,\nonumber\\
[E, d^*] = - d_c, \qquad [E, d_c^*] =  d, \qquad [F, d] =  d_c^*,
\qquad [F, d_c] =  - d^*.
\end{gather}
Thus $\mathcal K =  \mathfrak{sl}(2)  +\hspace{-3.9mm}\supset
 \mathfrak{hei}(0|4)$, where
 $\mathfrak{hei}(0|4)$ is the Heisenberg Lie superalgebra:
 $\mathfrak{hei}(0|4)_{\bar{1}} =\langle d, d_c^*\rangle
\oplus\langle d_c, d^*\rangle $
 is a direct sum  of two isotropic subspaces
 with respect to the non-degenerate symmetric form
 given by $(d, d^*) = (d_c, d_c^*) = 1$, and
 $\mathfrak{hei}(0|4)_{\bar{0}} = \langle \triangle\rangle $ is
the center.
 The isotropic subspaces are
standard $\mathfrak{sl}(2)$-modules.
Since $\mathfrak{sl}(2)\simeq \mathfrak{sp}(2)$, the following is
a natural generalization.

\subsection{Hyper-K\"ahler manifolds}

Let $M$ be a compact hyper-K\"ahler manifold.  By definition, $M$
is a Riemannian
manifold endowed with three complex structures $I$, $J$, and $K$,
such that $I\circ J = -J\circ I = K$ and $M$ is
K\"ahler with respect to each of the complex structures $I$, $J$
and $K$.
Let $\omega_I$, $\omega_J$ and $\omega_K$ be the corresponding
closed 2-forms
on $M$.
Having fixed one of the complex structures, for example, $I$,
we obtain the Hodge theory as in the K\"ahler case.

For  each of the complex structures $I$, $J$ and $K$,
the operators $E_i$, $E_j$ and $E_k$ and their dual operators
$F_i$, $F_j$ and $F_k$
with respect to the
Hodge inner product are naturally defined.
One can also define differentials,
$d_c^l$ and  $(d_c^l)^*$, where $l = i, j, k$
for  $I$, $J$ and $K$.
Set
\begin{equation}
d_c^1 = d, \qquad (d_c^1)^* = d^*.
\end{equation}
Let $Q = \lbrace 1, i, j, k\rbrace$ be the set of indices
satisfying
the quaternionic identities.

In the  hyper-K\"ahler
case there is  a natural action of the Lie algebra
$\mathfrak{so}(5)\simeq\mathfrak{sp}(4)$ on
$H^*(M)$ [13]. The $\mathfrak{sp}(4)$ is spanned  by the operators
$E_i$, $F_i$, $K_i$, where $i$ runs through the set $Q \setminus
\lbrace 1\rbrace$,
and $H$.
The non-vanishing commutation relations are
\begin{gather}
[E_i, F_i] = H, \qquad [H, E_i] = 2E_i, \qquad [H, F_i] = -2F_i,
\qquad
[E_i, F_j] = K_{ij}, \nonumber\\
[K_i, K_j] = -2K_{ij},\qquad K_{ij} = - K_{ji},\qquad
[K_i, E_j] = -2E_{ij}, \qquad E_{ij} = - E_{ji},\nonumber\\
[K_i, F_j] = -2F_{ij}, \qquad F_{ij} = - F_{ji},
\qquad i\not  = j.
\end{gather}
Let $\mathcal H$ be a Lie superalgebra, whose even part is spanned
by the
$\mathfrak{sp}(4)$ and the Laplace operator $\triangle$, and the
odd part
is spanned by the differentials
$d_c^l$ and  $(d_c^l)^*$, where  $l\in Q$.
Thus $\dim\mathcal H = (11|8)$.
The non-vanishing commutation relations in $\mathcal H$
are (1.5) and the following relations (cf.~[2, 14, 15]):
\begin{gather}
[d_c^l, (d_c^l)^*] = \triangle, \qquad
[H, d_c^l] = d_c^l,\qquad [H,(d_c^l)^*] = -(d_c^l)^*, \nonumber\\
[E_i, (d_c^l)^*] = -d_c^{il},\qquad d_c^{-l} = - d_c^{l},\qquad
[F_i, d_c^l] = (d_c^{il})^*,\nonumber\\
 (d_c^{-l})^* = - (d_c^l)^*,\qquad
[K_i, d_c^l] = -d_c^{il}, \qquad [K_i, (d_c^l)^*] = -(d_c^{il})^*.
\end{gather}
Thus $\mathcal H = \mathfrak{sp}(4) +\hspace{-3.6mm}\supset
 \mathfrak{hei}(0|8)$, where
 $\mathfrak{hei}(0|8)$ is the   Heisenberg Lie  superalgebra:
 $\mathfrak{hei}(0|8)_{{\bar 1}} = \langle d_c^{l} , (d_c^{l})^*
|\ l \in Q\rangle$,
 $\mathfrak{hei}(0|8)_{{\bar 0}} = \langle \triangle\rangle $,
where
 $\langle \triangle\rangle $ is the center.
 $\mathfrak{hei}(0|8)_{{\bar 1}} = V_1 \oplus V_2$ is a~direct sum
of two
 isotropic subspaces with
 respect to the non-degenerate symmetric form:
 $(d_c^{a}, (d_c^{b})^*) = \delta_{ab}$ for $a, b \in Q$;
\begin{gather}
V_1 =\langle d_c^{1} + \sqrt{-2}d_c^{j} - d_c^{k}, \
 d_c^{1} - \sqrt{-2}d_c^{i} + d_c^{k}, \nonumber\\
\qquad \qquad (d_c^{i})^* + (d_c^{j})^* + \sqrt{-2}(d_c^{k})^*, \
 \sqrt{-2}(d_c^{1})^* + (d_c^{i})^* - (d_c^{j})^*\rangle
,\nonumber\\
V_2 =\langle d_c^{1} - \sqrt{-2}d_c^{i} - d_c^{k}, \
 d_c^{1} - \sqrt{-2}d_c^{j} + d_c^{k}, \nonumber\\
\qquad\qquad  (d_c^{i})^* - (d_c^{j})^* - \sqrt{-2}(d_c^{k})^*,\
\sqrt{-2}(d_c^{1})^* + (d_c^{i})^* + (d_c^{j})^*\rangle .
\end{gather}
The subspaces $V_1$ and $V_2$ are irreducible
 $\mathfrak{sp}(4)$-modules.

\section{Superconformal algebras}

A {\it superconformal algebra} (SCA  [8, 9]) is a complex
${\mathbb Z}$-graded Lie superalgebra $\mathfrak g =
\oplus_i\mathfrak g_i$,
such that
$\mathfrak g$ is simple,
$\mathfrak g$ contains the centerless Virasoro algebra,
i.e.\ the Witt algebra,
$L = \oplus_{n\in\mathbb Z}{\mathbb C}L_n$
with the commutation relations
$[L_m, L_n] = (m - n)L_{m+n}$
as a subalgebra, and
$ad L_0$ is diagonalizable with finite-dimensional eigenspaces:
$\mathfrak g_i = \lbrace x\in\mathfrak g \; |\;
 [L_0, x] = ix\rbrace,$
 so that  dim $\mathfrak g_i < C$,
where $C$ is a constant independent of~$i$.
(Other definitions of superconformal algebras,
embracing central extensions, are also popular, see~[4]; for an
intrinsic definition see~[6]).

In general, a SCA is spanned by a number of fields;
the Virasoro field is among them.
The basic example of a SCA is $W(N)$.
Let $\Lambda(N)$ be the
Grassmann algebra in $N$ variables
$\theta_1, \ldots, \theta_N$. Let
$\Lambda(1, N) =\mathbb C [t, t^{-1}]\otimes \Lambda (N)$ be a
supercommutative
superalgebra with natural multiplication and with the following
parity
of generators: $p(t) = \bar{0}$, $p(\theta_i) = \bar{1}$
for $i = 1, \ldots, N$.
By definition,
$W(N)$ is the Lie superalgebra of all derivations of
$\Lambda(1, N)$.
Let $\partial_t$ stand for $\frac{\partial}{\partial t}$ and
$\partial_i$ stand for $\frac{\partial}{\partial \theta_i}$.

The superalgebra $W(N)$ contains a one-parameter family of
SCAs $S'(N, \alpha)$.
By definition,
\begin{equation}
S(N, \alpha) = \lbrace D \in W(N) \mid
\mbox{Div}(t^{\alpha}D) = 0\rbrace \qquad \mbox{for} \ \ \alpha
\in {\mathbb C},
\end{equation}
where
$\mbox{Div}\left(f\partial_t + \sum\limits_{i=1}^N f_i
\partial_i\right) =
\partial_tf + \sum\limits_{i=1}^N (-1)^{p(f_i)}
\partial_i f_i$
for $f, f_i \in \Lambda (1, N)$.
The derived superalgebra $S'(N, \alpha) =  [S(N, \alpha), S(N,
\alpha)]$
is simple.

Let $\mathfrak g = S'(2, 1)$ or $W(4)$, respectively. Let
\begin{equation}
L_n = -t^{n+1}\partial_t -
\frac 12(n + 2)t^{n}\sum_{i=1}^N\theta_i\partial_i,
\end{equation}
and let $\mathfrak g_0$ be singled out by $L_0$.
There exist an isomorphisms $\varphi:\mathcal K\longrightarrow
S'(2, 1)_0$,
 and  a monomorphism $\psi:\mathcal H\longrightarrow W(4)_0$
in the case of a compact K\"ahler or hyper-K\"ahler manifold,
respectively.

\subsection{K\"ahler manifolds}

Let $N = 2$. $S'(2, 1)$ is spanned by 4 bosonic fields
$L_n$, $H_n$, $E_n$, $F_n$, where
$E_n$, $F_n$ and $H_n$ form the loop algebra of $\mathfrak
{sl}(2)$,
and 4 fermionic fields $X_n^i$, $Y_n^i$, $i = 1, 2$:
\begin{gather*}
H_n = t^n(\theta_1\partial_1 - \theta_2\partial_2),\qquad
E_n = t^n\theta_1\partial_2,\qquad F_n =
t^n\theta_2\partial_1,\nonumber
\end{gather*}
\begin{gather}
X_n^1 = t^n\theta_1\partial_t +
(n+1)t^{n-1}\theta_1\theta_2\partial_2,\qquad
X_n^2 = -t^{n+1}\partial_2,\nonumber\\
Y_n^1 = -t^{n+1}\partial_1,\qquad
Y_n^2 = t^n\theta_2\partial_t +
(n+1)t^{n-1}\theta_2\theta_1\partial_1.
\end{gather}
The commutation relations between $L_n$ and the fields, defined by
(2.3), are
\begin{gather}
[L_n, H_m] = -mH_{n+m}, \qquad [L_n, E_m] = -mE_{n+m}, \qquad
[L_n, F_m] = -mF_{n+m},\\
[L_n, X_m^i] = \left(\frac{n}{2} - m\right)X_{n+m}^i,\qquad
[L_n, Y_m^i] = \left(\frac{n}{2} - m\right)Y_{n+m}^i, \qquad  i =
1, 2.
\end{gather}
Clearly, the fields
$X_n^i$ and $Y_n^i$, where $i = 1, 2$, generate $S'(2, 1)$.

The Lie superalgebra $S'(2, 1)_0$ is isomorphic to the Lie
superalgebra
$\mathcal K$ of classical operators in K\"ahler geometry.
The isomorphism $\varphi$ is as follows:
\begin{gather}
\varphi(\triangle) = L_0,\qquad \varphi(H) = H_0,\qquad
\varphi(E) = E_0,\qquad \varphi(F) = F_0,\nonumber\\
\varphi(d) = X_0^1,\qquad \varphi(d^*) = Y_0^1,\qquad
\varphi(d_c) = X_0^2, \qquad \varphi(d_c^*) = Y_0^2.
\end{gather}

\subsection{Hyper-K\"ahler manifolds}

Let $N = 4$. The following 10 bosonic fields span a subalgebra of
$W(4)$ isomorphic to the loop algebra of $\mathfrak {sp}(4)$:
\begin{gather}
H_n = t^n(\theta_1\partial_1 + \theta_2\partial_2 -
\theta_3\partial_3 - \theta_4\partial_4),\qquad
E_n^i = t^n(\theta_1\partial_4 + \theta_2\partial_3),\nonumber\\
F_n^i = t^n(\theta_3\partial_2 + \theta_4\partial_1),\qquad
E_n^j = it^n(\theta_1\partial_3 + \theta_2\partial_4),\qquad
F_n^j = -it^n(\theta_3\partial_1 + \theta_4\partial_2),\nonumber\\
E_n^k = t^n(\theta_1\partial_3 - \theta_2\partial_4),\qquad
F_n^k = t^n(\theta_3\partial_1 - \theta_4\partial_2),\nonumber\\
K_n^i = it^n(\theta_1\partial_1 - \theta_2\partial_2 -
\theta_3\partial_3 + \theta_4\partial_4),\qquad
K_n^j = t^n(\theta_1\partial_2 - \theta_2\partial_1 +
\theta_3\partial_4 - \theta_4\partial_3),\nonumber\\
K_n^k = -it^n(\theta_1\partial_2 + \theta_2\partial_1 -
\theta_3\partial_4 - \theta_4\partial_3).
\end{gather}
Define 8 fermionic fields $X_n^l$, $Y_n^l$, where $l\in Q$.
Let
\begin{equation}
A_n^m = t^n\theta_m\partial_t +
(n+1)t^{n-1}\theta_m\sum_{i=1}^4\theta_i\partial_i, \qquad m = 1,
\ldots, 4.
\end{equation}
Let $\mathbf{X} = (x_{lm})$ and $\mathbf{Y} = (y_{lm})$
be the following complex
$4\times 4$ matrices, where
$l = 1, i, j, k$  and $m = 1, \ldots, 4$:
\begin{equation}
\mathbf{X} =
\left( \begin{array}{cccc}
1&-1&i&i\\
-i&-i&1&-1\\
1&1&-i&i\\
-i&i&-1&-1
\end{array}\right), \qquad
\mathbf{Y} =
\left( \begin{array}{cccc}
-1&1&i&i\\
-i&-i&-1&1\\
-1&-1&-i&i\\
-i&i&1&1
\end{array}\right).
\end{equation}
Set
\begin{gather}
X_n^l = \frac 12\sum_{m=1}^2x_{lm}A_n^m +
\frac{1}{2}\sum_{m=3}^4x_{lm}t^{n+1}\partial_m,\nonumber\\
Y_n^l = \frac{1}{2}\sum_{m=1}^2y_{lm}t^{n+1}\partial_m  +
\frac{1}{2}\sum_{m=3}^4y_{lm}A_n^m.
\end{gather}
The commutation relations between
$L_n$ and the fields, defined by
(2.7) and (2.10), are analogs of the relations
(2.4) and (2.5), respectively.

The zero modes of the Virasoro field $L_n$ and of the fields,
defined by (2.7)
and (2.10),
span a Lie superalgebra, which is isomorphic to the Lie
superalgebra
$\mathcal H$ of classical opera\-tors in hyper-K\"ahler geometry.

The monomorphism $\psi$ is given by the following formulas:
\begin{gather}
\psi(\triangle) = L_0, \qquad\psi(H) = H_0,\qquad
\psi(E_i) = E_0^i,\qquad \psi(F_i) = F_0^i,  \nonumber\\
\psi(K_i) = K_0^i, \qquad
\psi(d_c^l) = X_0^l,\qquad \psi((d_c^l)^*) = Y_0^l,
\end{gather}
where $i$ runs through the set $Q \setminus \lbrace 1\rbrace$, and
$l\in Q$.

\begin{statement}
The fields $X_n^l$ and $Y_n^l$, for all $l \in Q$, generate
$W(4)$.
\end{statement}

\section{Conclusion}

It is natural to expect that ``affinization''
of the classical operators
in the case of an infinite-dimensional manifold gives a SCA,
which should act on a relevant cohomology
complex. Recall that the Weil complex is used for definition of
the equivariant differential forms.
The infinite-dimensional generalization of the classical Weil
complex
is the semi-infinite Weil complex of a graded Lie algebra~[1].

{\samepage
In particular, let $\tilde{\mathfrak g} =
\mathfrak g\otimes \mathbb C [t, t^{-1}]$
be the loop algebra of a complex finite-dimensional Lie algebra
$\mathfrak g$.
Naturally, $\tilde{\mathfrak g} = \oplus_{n \in \mathbb
Z}\tilde{\mathfrak g}_n$.
Let
$\tilde{\mathfrak g}' = \oplus_{n \in \mathbb Z}\tilde{\mathfrak
g}_n'$
be the restricted dual of $\tilde{\mathfrak g}$.
The linear space $U=\tilde{\mathfrak g} \oplus \tilde{\mathfrak
g}'$
can be naturally endowed with non-degenerate skew-symmetric and
symmetric bilinear
forms: $(\cdot ,\cdot )$ and $\lbrace\cdot  , \cdot \rbrace$.
The  Weyl and Clifford algebras,
$W(\tilde{\mathfrak g})$ and~$C(\tilde{\mathfrak g})$, are the
quotients of the tensor algebra
$T^*(U)$
modulo the two-sided ideals generated by the elements of the form
$xy - yx - (x, y)$ and $xy + yx - \lbrace x,  y\rbrace$ for any
$x,y\in U$, respectively; here $xy:=x\otimes y$.
Let $u$ run through a fixed basis of ${\mathfrak g}$ and
$u'$ run through the dual basis.
Let
$\beta (u_m)$, $\gamma (u_m')$ and $\tau (u_m)$, $\varepsilon
(u_m')$,
where $m\in\mathbb Z$, be
generators of $W(\tilde{\mathfrak g})$ and $C(\tilde{\mathfrak
g})$,
respectively.
We can realize $S'(2, 1)$ in terms of the following quadratic
expansions:
\begin{gather}
L_n = \sum_{u, m}  m : \tau (u_{m-n})\varepsilon (u_m') :
+ m : \beta (u_{m-n})\gamma (u_m') : -
\frac{n}{2}  : \beta (u_m) \gamma (u_{m+n}') :\ ,\nonumber\\
H_n = - \sum_{u, m} : \beta (u_m) \gamma (u_{m+n}') :\ ,\qquad
E_n = - \frac{i}{2} \sum_{u, m}
\gamma(u_m')\gamma(u_{n-m}'),\nonumber\\
F_n = - \frac{i}{2} \sum_{u, m} \beta(u_m)\beta(u_{-m-n}),\qquad
X_n^1 = \sum_{u, m} \gamma (u_{m+n}')\tau (u_m), \nonumber\\
X_n^2  = i\sum_{u, m}  m\gamma (u_{m-n}')\varepsilon (u_m'),\qquad
Y_n^1 = \sum_{u, m} m\beta (u_{m-n})\varepsilon (u_m'),
\nonumber\\
Y_n^2  = i\sum_{u, m} \beta (u_m)\tau (u_{-m-n}),
\end{gather}
where the double colons
$:\mbox { } :$ denote a normal ordering operation:
\[
: \tau (u_j)\varepsilon (v_i') : =
\left\{
\begin{array}{ll}
\tau (u_j)\varepsilon (v_i') & \mbox{if} \ \ i \leq 0,\\
- \varepsilon (v_i')\tau (u_j) & \mbox {if } \ \ i > 0
\end{array}\right.
\]
with the similar formula for $\beta$ and $\gamma$, but without
the minus sign.}

The semi-infinite Weil complex of $\tilde{\mathfrak g}$ is
\begin{equation}
\left\{ S^{\frac{\infty}{2} + *}(\tilde{\mathfrak g})\otimes
\Lambda^{\frac{\infty}{2} + *}(\tilde{\mathfrak g}), \ \
\mbox{\bf d} + \mbox{\bf h} \right\},
\end{equation}
where $S^{\frac{\infty}{2} + *}(\tilde{\mathfrak g})$ and
$\Lambda^{\frac{\infty}{2} + *}(\tilde{\mathfrak g})$ are
semi-infinite
analogs of the modules of symmetric and exterior powers (see~[1]),
$\mbox{\bf d}$ is the analog of the differential for Lie algebra
(co)homology and $\mbox{\bf h}$ is the analog of the Koszul
differential:
\begin{gather}
\mbox{\bf d}   = \sum_{u,v,i,j} \frac{1}{2}: \tau ([u_i,
v_j])\varepsilon (v_j')
\varepsilon (u_i') : +
: \beta ([u_i, v_j])\gamma (v_j')  \varepsilon (u_i') :\ ,
\nonumber\\
\mbox{\bf h}  = \sum_{u,i} \gamma (u_i')\tau (u_i),
\end{gather}
The quadratic operators (3.1) define a projective action of $S'(2,
1)$
on the semi-infinite Weil complex of $\tilde{\mathfrak g}$. The
cocycle
is (see~[8])
\begin{gather}
c(L_n, L_k) = \frac{n^3}{12}\delta_{n, -k},\qquad
c(E_n, F_k) = \frac{n-1}{6}\delta_{n, -k}, \qquad
c(L_n, H_k) = -\frac{n}{6}\delta_{n, -k}, \nonumber\\
c(H_n, H_k) =  \frac{n}{3}\delta_{n, -k},\qquad
 c(X_n^i, Y_k^i) =   \frac{n(n-1)}{6}\delta_{n, -k}, \qquad i = 1,
2.
\end{gather}
If ${\mathfrak g}$ has a non-degenerate invariant symmetric
bilinear form,
then this action commutes with $\mbox{\bf d}$, and the action on
the
(relative) semi-infinite cohomology is well-defined (see~[12]).
The restriction of this action to the zero modes defines a
representation
of~$\mathcal K$.
Note that in this way $d$ and $d^*$ act as
the semi-infinite Koszul differential $\mbox{\bf h}$ and
the semi-infinite homotopy operator, respectively; this is a
generalization of
Howe's construction~[7]. Observe that our superalgebras
$\mathcal K$ and $\mathcal H$ differ from the ones usually
considered in examples of Howe duality on (hyper)K\"ahler
manifolds
but are contractions of $\mathfrak{osp}(2|2)$ and
   $\mathfrak{osp}(2|4)$, cf.~[11].

It was shown in~[3] that the relative semi-infinite complex of a
 $\mathbb Z$-graded complex
Lie algebra with coefficients in a graded Hermitian module has
a structure similar to that of the de Rham complex in
K\"ahler geometry.
In~[12] we described operators on the relative semi-infinite Weil
complex
of the loop algebra of a complex Lie algebra, which are analogs of
the
classical ones in
K\"ahler geometry
 and span a Lie superalgebra isomorphic to
$\mathcal K$. Note that in this realization (for which  no
``affinization'' seems to be possible) $d$ acts as the
differential
$\mbox{\bf d}$.

An interesting problem is to define operators acting on a
(relative) semi-infinite Weil complex, which are analogous to the
classical ones
in hyper-K\"ahler geometry, and obtain the corresponding
field expansions.

\subsection*{Acknowledgements}

This work is supported by the Anna-Greta and Holder Crafoords
fond,
The Royal Swedish Academy of Sciences.
This work was started at the Max-Planck-Institut f\"ur
Mathematik, Bonn.
I wish to thank MPI for the hospitality and support.

\label{poletaeva-lastpage}

\end{document}